\documentclass[11pt]{article}
\usepackage{graphicx}
\usepackage{ifpdf}
\usepackage[letterpaper, hmargin=1in, vmargin=1.25in]{geometry}

\begin{document}

 \title{Adleman-Manders-Miller Root Extraction Method Revisited}

 \author{Zhengjun Cao\,$^*$, \quad Qian Sha,\quad  Xiao Fan\\
  Department of Mathematics, Shanghai University, Shanghai,
  China.\\
  \textsf{$^*$\ \textsf{caozhj@shu.edu.cn}}  }

 \date{}\maketitle

\maketitle

\begin{abstract} In 1977, Adleman, Manders and Miller
 had briefly described  how to
extend their square root extraction method to the general $r$th root
extraction over finite fields, but  not shown enough details.
Actually, there is a dramatic difference between the square root extraction and
the general $r$th root extraction because one has to solve discrete
logarithms for  $r$th root extraction. In this paper, we clarify
their method and analyze its complexity. Our heuristic presentation
is helpful to grasp the method entirely and deeply.

  \end{abstract}


\section{Introduction}

Root extraction is a classical problem in computers algebra. It
is essential to cryptosystems based on elliptic curves
\cite{BV06}. There are several efficient probabilistic algorithms
for square
  root extraction in finite fields, such as Cipolla-Lehmer
  \cite{CI03,Le69},
  Tonelli-Shanks \cite{Sh72,To91}
 and Adleman-Manders-Miller
\cite{AMM77}. All of them require a quadratic nonresidue as an
additional input. In 2004, M\"{u}ller investigated this topic in
Ref.\cite{M04}. In 2011, Sze \cite{S11}  presented a novel idea to
compute square roots over finite fields, without being given any
quadratic nonresidue, and without assuming any unproven hypothesis.

 Adleman-Manders-Miller square root extraction method can
be extended to solve the general  $r$th root extraction problem.
  In recent,  Nishihara et al. \cite{NHSK}
have specified the Adleman-Manders-Miller method for cube root
extraction.
 Barreto and Voloch \cite{BV06} proposed an efficient algorithm to
compute $r$th roots in $ {F}_{p^m} $ for certain choices of
$m$ and $p$. Besides, it requires that  $r\,||\,p-1$ and $(m, r)=1 $, where the
notation $a^b||c$ means that $a^b$ is the highest power of $a$
dividing $c$.

The basic idea of Adleman-Manders-Miller square root extraction
 in $ {F}_{p} $ can be described as follows. Write $p-1$
in the form $2^{t}\cdot s$, where $s$ is odd.  Given a quadratic
residue $\delta$ and a quadratic nonresidue $\rho$, we have
$$\left(\delta^s\right)^{2^{t-1}}\equiv 1 \ (\mbox{mod}\ p), \quad
 \left(\rho^s\right)^{2^{t-1}}\equiv -1 \ (\mbox{mod}\ p) $$
If  $t\geq 2$, then  $ \left(\delta^s\right)^{2^{t-2}} \
(\mbox{mod}\ p) \in\{1, -1\} $.
 Take $k_1=0$ or $1$ such that
$$\left(\delta^s\right)^{2^{t-2}}
\left(\rho^s\right)^{2^{t-1}\cdot k_1}\equiv 1  \ (\mbox{mod}\ p)
$$
Since $
\left(\delta^s\right)^{2^{t-3}}\left(\rho^s\right)^{2^{t-2}\cdot
k_1}
 \ (\mbox{mod}\ p)
\in\{1, -1\}$, take $k_2=0$ or $1$ such that
$$\left(\delta^s\right)^{2^{t-3}}\left(\rho^s\right)^{2^{t-2}\cdot k_1}
\left(\rho^s\right)^{2^{t-1}\cdot k_2}\equiv 1  \ (\mbox{mod}\ p)
$$
Likewise, we can obtain $k_3, \cdots, k_{t-1}\in \{0, 1\}$ such that
$$\left(\delta^s\right)\left(\rho^s\right)^{2\cdot k_1
+2^{2}\cdot k_2+\cdots +2^{t-1}\cdot k_{t-1} }\equiv 1 \
(\mbox{mod}\ p)
$$
Thus, we have
$$ \left(\delta^{\frac{s+1}2}\right)^2\left(\left(\rho^s\right)^{k_1
+2\cdot k_2+\cdots +2^{t-2}\cdot k_{t-1} }\right)^2\equiv \delta \
(\mbox{mod}\ p) $$

It should be stressed, however, that  there is a dramatic difference  between the
square root extraction and the general $r$th root extraction. Write
$p-1$ in the form $r^{t}\cdot s$, where $(r, s)=1$. Given a $r$th
residue $\delta$ and a $r$th nonresidue $\rho$, we have
$$\left(\delta^s\right)^{r^{t-1}}\equiv 1 \ (\mbox{mod}\ p), \quad
 \left(\rho^s\right)^{r^{t-1}}\not\equiv 1 \ (\mbox{mod}\ p) $$
 Since    $ \left(\delta^{s}\right)^{r^{t-2}} \
(\mbox{mod}\ p)$  is a root of the equation $X^r\equiv 1 \
(\mbox{mod}\ p)$ and the equation has $r$ different roots (these
roots can be represented by $\left(\rho^s\right)^{k_i\cdot
r^{t-1}},\, k_i\in \{0, 1, \cdots, r-1 \} $), it becomes difficult
to find
   $k_1$  such that
$$\left(\delta^{s}\right)^{r^{t-2}}
\left(\rho^s\right)^{r^{t-1}\cdot k_1}\equiv 1  \ (\mbox{mod}\ p)
$$

In 1977,  Adleman, Manders and Miller \cite{AMM77}
had  presented a brief description on how to
extend their square root extraction method to the general $r$th root
extraction over finite fields, but not shown enough details.
By the way, it is the only known method for the
general $r$th root extraction over finite fields.  In this paper, we clarify their  method and analyze
its complexity.

\section{Preliminary}

Let $ {Z}_n=\{0, 1, \cdots, n-1\} $ be the set of all numbers
smaller than $n$, $ {Z}_n^*=\{x\,|\,1\leq x\leq n \
\mbox{and}\  \mbox{gcd}(x, n)=1 \} $ be the set of numbers in
$ {Z}_n $ that are coprime to $n$. The following definitions
and results can be found in Ref.\cite{LN}.

Definition 1. \emph{A residue $a\in  {Z}_n^*$ is said to be a quadratic residue
 if there exists some $x\in  {Z}_n^*$ such that
 $x^2\equiv a \ (\emph{\mbox{mod}}\ n) $.
 If $a$ is not a quadratic residue, then it is referred to as a quadratic non-residue.}

Theorem 2. \emph{\emph{(Euler's Criterion)} For prime $p$,
an element $a\in  {Z}_p^*$ is a quadratic residue if and only
if $a^{\frac{p-1}2}\equiv 1  \ (\emph{\mbox{mod}}\ p)$.}

Definition 3.  \emph{\emph{(Legendre Symbol)} For any prime
$p$ and $a\in  {Z}_p^*$, we define the Legendre symbol
$$\left[\frac a p\right] = \left\{
\begin{array}{ll}
1 & \emph{\mbox{if $a$ is a quadratic residue}}  \ (\emph{\mbox{mod}}\ p)\\
-1 & \emph{\mbox{if $a$ is a quadratic non-residue}}\ (\emph{\mbox{mod}}\ p)\\
\end{array}
\right.
$$
}

For an integer $a$, we define log$(a)$ to be the number of bits in
the binary representation of $|a|$; more precisely, $$ \mbox{log}(a)
=\left\{
\begin{array}{ll}
 \lfloor\log_2 |a|\rfloor+1 & \mbox{if}\  a\neq 0 \\
1 & \mbox{if}\  a= 0 \\
\end{array}
\right. $$

Given  $a\in  {Z}_n$ and a non-negative integer $e$, the
repeated-squaring algorithm  computes   $a^e  \ (\mbox{mod}\ n)$
using just $\mathcal {O}(\mbox{log}(e))$ multiplications in
$ {Z}_n$, thus taking time $\mathcal
{O}(\mbox{log}(e)\mbox{log}^2n)$. Therefore, we have the following
result:

Proposition 4.  \emph{For an odd prime $p$, we can test whether an integer $a$
  is a quadratic
residue modulo $p$ by either performing the exponentiation
$a^{\frac{(p-1)}{2}}  \ (\emph{\mbox{mod}}\ p)$ or by computing the
Legendre symbol $\left[\frac a p\right]$. Assume that $0<
  a < p$. Using a standard repeated squaring algorithm, the former
method takes time $ \mathcal {O}( \emph{\mbox{log}}^3p)$, while
using Euclidean-like algorithm, the latter method takes time
$\mathcal {O}(\emph{\mbox{log}}^2p)$.}

\emph{Proof}. See \cite{S05}.

Let $R$ be a ring.  Let us define the length of a polynomial
$f(X)\in R[X]$, denoted by log$(f)$, to be the length of its
coefficient vector; more precisely, we define
$$ \mbox{log}(f)
=\left\{
\begin{array}{ll}
 \deg(f)+1 & \mbox{if}\  f\neq 0 \\
1 & \mbox{if}\  f= 0 \\
\end{array}
\right. $$

Analogous to algorithms for modular integer arithmetic, we can also
do arithmetic in the residue class ring $R[X]/(f)$, where $f \in
R[X]$ is a polynomial of $\deg(f)>0$ whose leading coefficient
lc$(f)$ is a unit.

Proposition 5.  \emph{Let  $R[X]/(f)$ be a residue class ring,
 where $f \in R[X]$ is a polynomial of $\deg(f)>0$ whose leading
coefficient \emph{lc}$(f)$ is a unit.
  Given $g \in R[X]/(f)$ and a non-negative
exponent $e$,  using repeated-squaring algorithm  we can compute
$g^e$ taking $\mathcal {O}(\emph{\mbox{log}}(e)\deg(f)^2)$
operations in $R$.}

\emph{Proof}. See \cite{GG03}.  

Notice that using a standard representation for $ {F}_p $,
each operation in $ {F}_p $ takes time $\mathcal
{O}(\emph{log}^2p)$.

\section{Adleman-Manders-Miller square root extraction method }

The Adleman-Manders-Miller square root extraction method requires a
quadratic non-residue as an additional input. We classify the
 method into two kinds because there is a gap between the base field $ {F}_p $ and
 the extension $ {F}_{p^m} $ to test whether an element
 is a quadratic non-residue.

\subsection{Adleman-Manders-Miller square root extraction method in $ {F}_p $}

Consider the problem to find a solution to the congruence $X^2\equiv
\delta \ (\mbox{mod}\ p) $ over finite field $ {F}_p $, where
$p$ is an odd prime.

  Adleman, Manders and Miller \cite{AMM77} proposed an algorithm to solve the
  problem. Their square root extraction method is based on the
following facts. Write $p-1$ in the form $2^{t}\cdot s$, where $s$
is odd.  Given a quadratic residue $\delta$ and a quadratic
nonresidue $\rho$, we have
$$\left(\delta^s\right)^{2^{t-1}}\equiv 1 \ (\mbox{mod}\ p), \quad
 \left(\rho^s\right)^{2^{t-1}}\equiv -1 \ (\mbox{mod}\ p) $$
If $t=1$, then   $\delta^s\equiv 1 \ (\mbox{mod}\ p)$.
  Hence, we have $\left(\delta^{\frac{s+1}2}\right)^2\equiv \delta \
(\mbox{mod}\ p) $.  It means that $\delta^{\frac{s+1}2}$ is a square
root of $\delta$. In this case, it only takes time $\mathcal
{O}(\mbox{log}(s)\mbox{log}^2p)$.

If $t\geq 2$, then  $ \left(\delta^s\right)^{2^{t-2}} \ (\mbox{mod}\
p) \in\{1, -1\} $.
 Take $k_1=0$ or $1$ such that
$$\left(\delta^s\right)^{2^{t-2}}
\left(\rho^s\right)^{2^{t-1}\cdot k_1}\equiv 1  \ (\mbox{mod}\ p)
$$
 Take $k_2=0$ or $1$ such that
$$\left(\delta^s\right)^{2^{t-3}}\left(\rho^s\right)^{2^{t-2}\cdot k_1}
\left(\rho^s\right)^{2^{t-1}\cdot k_2}\equiv 1  \ (\mbox{mod}\ p)
$$
Likewise, we obtain $k_3, \cdots, k_{t-1}\in \{0, 1\}$ such that
$$\left(\delta^s\right)\left(\rho^s\right)^{2\cdot k_1
+2^{2}\cdot k_2+\cdots +2^{t-1}\cdot k_{t-1} }\equiv 1 \
(\mbox{mod}\ p)
$$
Finally, we have
$$ \left(\delta^{\frac{s+1}2}\right)^2\left(\left(\rho^s\right)^{k_1
+2\cdot k_2+\cdots +2^{t-2}\cdot k_{t-1} }\right)^2\equiv \delta \
(\mbox{mod}\ p) $$

\begin{table}[!h]
\tabcolsep 0pt
\caption{Adleman-Manders-Miller square root extraction algorithm in
$ {F}_p $}
\vspace*{-12pt}
\begin{center}
\def\temptablewidth{1\textwidth}
{\rule{\temptablewidth}{.6pt}}

\begin{tabular*}{\temptablewidth}{@{\extracolsep{\fill}}l}
 {Input}: Odd prime $p$ and a quadratic residue
$ \delta.$\\
{Output}:  A square root  of  $ \delta$.
  \\ \hline
{Step 1:} Choose $\rho$ uniformly at random from $ {F}_p^* $.\\
\hspace*{11mm}Compute $[\frac{\rho}p]$
  using Euclidean-like algorithm.\\
 {Step 2:} \textbf{if} $[\frac{\rho}p]=1$, \textbf{go to} Step
 1.\\
{Step 3:} Compute $t, s  $ such that $p-1=2^{t}
s $, where $s$ is odd.\\
\hspace*{11mm}Compute  $ a\leftarrow \rho^{s}, b\leftarrow
\delta^{s},$ $h\leftarrow 1 $.
\\ {Step 4:}
\textbf{for} $i=1$ \textbf{to} $t-1$ \\
\hspace*{16mm}compute $d=b^{2^{t-1-i}}$\\
\hspace*{16mm}\textbf{if} $d= 1$,  $k\leftarrow 0$\\
 \hspace*{16mm}\textbf{else} $k\leftarrow 1$\\
 \hspace*{16mm}$b\leftarrow b\cdot (a^2)^{k}$, $h\leftarrow h \cdot
a^{k}$ \\
\hspace*{16mm}$a\leftarrow a^2$\\
 \hspace*{11mm}\textbf{end for}
 \\
{Step 5:} \textbf{return} $\delta^{\frac{s+1}2}\cdot h $
 \end{tabular*}
       {\rule{\temptablewidth}{.6pt}}
       \end{center}
       \end{table}

To find a quadratic non-residue  $\rho$, it requires to check that
 $[\frac{\rho}p]\neq 1$. The computation takes time
$\mathcal {O}(\mbox{log}^2p)$.
 If we do this for more than
$\mathcal {O}(1)\mbox{log}\,p  $ different randomly chosen $\rho$,
then with probability $>1-(\frac 1 p)^{\mathcal {O}(1)}$ at least
one of them will give a quadratic non-residue. Thus, to find a
quadratic nonresidue $\rho$, it takes expected time $\mathcal
{O}(\mbox{log}^3p)$. To compute $ b^{2^{t-i-1}}\ (\mbox{mod}\ p) $,
it takes time $\mathcal {O}((t-i-1)\mbox{log}^2p)$. Since there are
$1+2+\cdots +(t-1)=\frac{t(t-1)}2$ steps, the
   loop takes time  $\mathcal {O}(t^2\mbox{log}^2p)$.
Thus, the total estimate is $\mathcal
{O}(\mbox{log}^3p+t^2\mbox{log}^2p)$. At worst (if almost all of
$p-1$ is a power of 2), this is $\mathcal {O}(\mbox{log}^4p)$.

\subsection{Adleman-Manders-Miller square root extraction method in $ {F}_{p^m} $}

As we mentioned before, the Adleman-Manders-Miller method in the
extension field $ {F}_{p^m} $ differs from the method in the
base field $ {F}_{p} $ because one can not determine a
quadratic non-residue by computing the Legendre Symbol.

Set $q=p^m$. To find a quadratic non-residue  $\rho$, it requires to
check that
 $\rho^{\frac{q-1}2}\neq 1$. The computation takes time
$\mathcal {O}(\mbox{log}^3q)$.
 If we do this for more than
$\mathcal {O}(1)\mbox{log}\,q  $ different randomly chosen $\rho$,
then with probability $>1-(\frac 1 q)^{\mathcal {O}(1)}$ at least
one of them will give a quadratic non-residue. Thus, to find a
quadratic nonresidue $\rho$, it takes expected time $\mathcal
{O}(\mbox{log}^4q)$.

To compute $ b^{2^{t-i-1}} $, it takes time $\mathcal
{O}((t-i-1)\mbox{log}^2q)$. Since there are  $1+2+\cdots +(t-1)$
steps, the
   loop takes time  $\mathcal {O}(t^2\mbox{log}^2q)$.
Thus, the final estimate is $\mathcal
{O}(\mbox{log}^4q+t^2\mbox{log}^2q)$.

\begin{table}[!h]
\tabcolsep 0pt
\caption{Adleman-Manders-Miller square root extraction algorithm in
$ {F}_{p^m} $ }
\vspace*{-12pt}
\begin{center}
\def\temptablewidth{1\textwidth}
{\rule{\temptablewidth}{.6pt}}

\begin{tabular*}{\temptablewidth}{@{\extracolsep{\fill}}l}

{Input}: Odd prime $p$, a positive integer $m$ and a quadratic
residue
$ \delta.$ \\
{Output}:  A square root  of  $ \delta$.
  \\ \hline
{Step 1:} Choose $\rho$ uniformly at random from $ {F}_{p^m}^* $.\\
 {Step 2:} \textbf{if} $\rho^{\frac{p^m-1}2}=1$, \textbf{go to} Step
 1.\\
{Step 3:} Compute $t, s  $ such that $p^m-1=2^{t}
s $, where $s$ is odd.\\
\hspace*{11mm}Compute  $ a\leftarrow \rho^{s}, b\leftarrow
\delta^{s},$ $h\leftarrow 1 $.
\\ {Step 4:}
\textbf{for} $i=1$ \textbf{to} $t-1$ \\
\hspace*{16mm}compute $d=b^{2^{t-1-i}}$\\
\hspace*{16mm}\textbf{if} $d= 1$,  $k\leftarrow 0$\\
 \hspace*{16mm}\textbf{else} $k\leftarrow 1$\\
 \hspace*{16mm}$b\leftarrow b\cdot (a^2)^{k}$, $h\leftarrow h \cdot
a^{k}$ \\
\hspace*{16mm}$a\leftarrow a^2$\\
 \hspace*{11mm}\textbf{end for}
 \\
{Step 5:} \textbf{return} $\delta^{\frac{s+1}2}\cdot h $\end{tabular*}
       {\rule{\temptablewidth}{.6pt}}
       \end{center}
       \end{table}

\section{Adleman-Manders-Miller cubic root extraction method}

 In 2009,  Nishihara et al. \cite{NHSK}  specified
the Adleman-Manders-Miller method for cube root extraction. See the
following description.

\begin{table}[!h]
\tabcolsep 0pt
\caption{Adleman-Manders-Miller cubic root extraction algorithm in
$ {F}_{p^m} $}
\vspace*{-12pt}
\begin{center}
\def\temptablewidth{1\textwidth}
{\rule{\temptablewidth}{.6pt}}

\begin{tabular*}{\temptablewidth}{@{\extracolsep{\fill}}l}

{Input}: Odd prime $p$, a positive integer $m$ and a cubic residue
$ \delta.$ \\
{Output}:  A cubi root  of  $ \delta$.
  \\ \hline
{Step 1:} Choose $\rho$ uniformly at random from $ {F}_{p^m}^* $.\\
 {Step 2:} \textbf{if} $\rho^{\frac{p^m-1}3}=1$, \textbf{go to} Step
 1.\\
{Step 3:} Compute $t, s  $ such that $p^m-1=3^{t}
s $, where $s=3l\pm 1$.\\
\hspace*{11mm}Compute  $ a\leftarrow \rho^{s}, a'\leftarrow
\rho^{3^{t-1}\cdot s}, b\leftarrow \delta^{s},$ $h\leftarrow 1 $.
\\ {Step 4:}
\textbf{for} $i=1$ \textbf{to} $t-1$ \\
\hspace*{16mm}compute $d=b^{3^{t-1-i}}$\\
\hspace*{16mm}\textbf{if} $d= 1$,  $k\leftarrow 0$,\\
 \hspace*{16mm}\textbf{else if} $d=a' $,  $k\leftarrow 2$\\
 \hspace*{16mm}\textbf{else}  $k\leftarrow 1$\\
 \hspace*{16mm}$b\leftarrow b\cdot (a^3)^{k}$, $h\leftarrow h \cdot
a^{k}$ \\
\hspace*{16mm}$a\leftarrow a^3$\\
 \hspace*{11mm}\textbf{end for}
 \\
{Step 5:}\ \ \ $r\leftarrow \delta^l h $\\
 \hspace*{13mm}\textbf{if} $s=3l+1$, $r\leftarrow r^{-1}$\\
\hspace*{13mm}\textbf{return} $r$\end{tabular*}
       {\rule{\temptablewidth}{.6pt}}
       \end{center}
       \end{table}

Set $q=p^m$.   The cubic root extraction algorithm takes time
$\mathcal {O}(\mbox{log}^4q+t^2\mbox{log}^2q)$. As for this claim,
we refer to  the  complexity analysis of Adleman-Manders-Miller
square root extraction algorithm in Section 3.2.

\section{Specification of Adleman-Manders-Miller $r$th Root Extraction Method}

Consider the general problem to find a solution to  $X^r= \delta  $
in $ {F}_{q} $.
Clearly, it suffices to consider the following two cases:\\
\centerline {(1) $(r, q-1)=1$; \qquad  (2) $r| q-1 $.}

If $(r, q-1)=1$, then $\delta^{r^{-1}}$ is a $r$th root of $\delta$.
Therefore, it suffices to consider the case that $r| q-1 $.

Adleman, Manders and Miller \cite{AMM77} had mentioned how to
extend their square root extraction method to   $r$th root
extraction, but  not specified it.  We now clarify it as
follows.

 If $r|q-1$, we write $p-1$ in the form $r^{t}\cdot s$, where $(s, r)=1$.
 Given a $r$th residue $\delta$, we have
$\left(\delta^{s}\right)^{r^{t-1}}= 1   $. Since $(s, r)=1$,  it is
easy to find the least nonnegative integer $\alpha $ such that
$s|r\alpha -1$. Hence,
$$\left(\delta^{r\alpha -1}\right)^{r^{t-1}}= 1   \eqno(1)$$
If $t-1=0$, then $\delta^{\alpha }$ is a $r$th root of $\delta$.
From now on, we assume that $t \geq 2$.

 Given a $r$th non-residue $\rho\in  {F}_{q}$,  we have
 $$\left(\rho^s\right)^{i\cdot r^{t-1}} \neq \left(\rho^s\right)^{j\cdot r^{t-1}}
 \ \mbox{where} \ i\neq j,\  i, j\in \{0, 1, \cdots, r-1\} $$
 Set
$$K_i=\left(\rho^s\right)^{i\cdot r^{t-1}} \  \mbox{and}\    {K}=\{K_0,
K_1, \cdots,   K_{r-1}\}$$ It is easy to find that all $K_i$ satisfy
$X^r = 1   $. Since $$ \left(\left(\delta^{r\alpha
-1}\right)^{r^{t-2}}\right)^r = 1  $$ there is a unique $j_1\in \{0,
1, \cdots, r-1\}$ such that
$$\left(\delta^{r\alpha -1}\right)^{r^{t-2}}=K_{r-j_1}   $$
where $K_r=K_0$.
 Hence,
$$\left(\delta^{r\alpha -1}\right)^{r^{t-2}}K_{j_1}= 1  $$ That is
$$\left(\delta^{r\alpha -1}\right)^{r^{t-2}}\left(\rho^s\right)^{j_1\cdot r^{t-1}}
= 1  \eqno(2) $$
By the way, to obtain $j_1$ one has to solve a discrete logarithm.

Likewise, there is a  unique $j_2 \in \{0, 1,
\cdots, r-1\}$ such that
$$\left(\delta^{r\alpha -1}\right)^{r^{t-3}}\left(\rho^s\right)^{j_1\cdot
r^{t-2}}\left(\rho^s\right)^{j_2\cdot r^{t-1}} = 1  \
 \eqno(3)$$ Consequently, we can obtain $j_1, \cdots, j_{t-1} $
such that
$$\left(\delta^{r\alpha -1}\right)\left(\rho^s\right)^{j_1\cdot
r}\left(\rho^s\right)^{j_2\cdot r^2}\cdots
\left(\rho^s\right)^{j_{t-1}\cdot r^{t-1}} = 1   \eqno(4)$$
 Thus, we have
 $$\left(\delta^{\alpha}\right)^r\left(\left(\rho^s\right)^{j_1 + j_2\cdot r
 +\cdots j_{t-1}\cdot r^{t-2}}\right)^r
= \delta   \eqno(5)$$ It means that $$\delta^{\alpha}
\left(\rho^s\right)^{j_1 + j_2\cdot r
 +\cdots j_{t-1}\cdot r^{t-2}} $$ is a $r$th root of
 $\delta$.

 \begin{table}[!h]
\tabcolsep 0pt
\caption{Adleman-Manders-Miller $r$th root extraction algorithm in
$ {F}_{q} $}
\vspace*{-12pt}
\begin{center}
\def\temptablewidth{1\textwidth}
{\rule{\temptablewidth}{.6pt}}

\begin{tabular*}{\temptablewidth}{@{\extracolsep{\fill}}l}

{Input}: $ {F}_{q} $ and a $r$th residue
$ \delta,$ $r|q-1$. \\
{Output}:  A $r$th root of $\delta$.
  \\ \hline
{Step 1:} Choose $\rho$ uniformly at random from $ {F}_q^* $.\\
 {Step 2:} \textbf{if} $\rho^{\frac{q-1}r}= 1$, \textbf{go to} Step
 1.\\
{Step 3:} Compute $t, s  $ such that $q-1=r^{t}
s $, where $(r, s)=1$.\\
\hspace*{11mm}Compute the least nonnegative integer $\alpha $ such
that $s|r\alpha -1$.\\  \hspace*{11mm}Compute  $ a\leftarrow
\rho^{r^{t-1}s}, b\leftarrow \delta^{r\alpha -1}$, $c\leftarrow
\rho^s$, $h\leftarrow 1 $
\\ {Step 4:}
\textbf{for} $i=1$ \textbf{to} $t-1$ \\
\hspace*{16mm}compute $d=b^{r^{t-1-i}}$\\
\hspace*{16mm}\textbf{if} $d=1$,  $j\leftarrow 0$,\\
\hspace*{16mm}\textbf{else} $j\leftarrow -\log_{a}d$ (compute the discrete logarithm)\\
\hspace*{16mm}$b\leftarrow b\,(c^r)^j $, $h\leftarrow h\, c^{j}$\\
\hspace*{16mm}$c\leftarrow c^{r} $\\
 \hspace*{11mm}\textbf{end
for}
 \\
{Step 5:} \textbf{return} $\delta^{\alpha}\cdot h $\end{tabular*}
       {\rule{\temptablewidth}{.6pt}}
       \end{center}
       \end{table}

 \section{Complexity analysis of Adleman-Manders-Miller $r$th Root Extraction Method}

We now discuss the time estimate for this $r$th root extraction
algorithm.

To find a $r$th non-residue  $\rho$, it requires to check that
$\rho^{\frac{q-1}r}\neq 1$. The computation takes time $\mathcal
{O}(\mbox{log}^3q)$.
 If we do this for more than
$\mathcal {O}(1)\mbox{log}\,q  $ different randomly chosen $\rho$,
then with probability $>1-(\frac 1 q)^{\mathcal {O}(1)}$ at least
one of them will give a $r$th non-residue. Therefore,  the expected time
of finding a $r$th non-residue  is $\mathcal {O}(\mbox{log}^4q)$.

The work done outside the loop
 amounts to just a handful of exponentiations.
  Hence, it takes time $\mathcal {O}(\mbox{log}^3q)$.
 To compute $ b^{r^{t-i-1}} $, it takes time $\mathcal
{O}((t-i-1)\mbox{log}\, r\mbox{log}^2q)$. Since there are
$1+2+\cdots +(t-1)$ steps, it takes time  $\mathcal
{O}(t^2\mbox{log}\,r\mbox{log}^2q)$.

 To compute the discrete
logarithm $\log_{a}d $, it takes time $\mathcal {O}(r
\mbox{log}^2q)$ using brute-force search. Since there are $t-1$
discrete logarithms at worst, it takes time $\mathcal
{O}(tr\mbox{log}^2q)$.

Thus, the final estimate is $\mathcal
{O}(\mbox{log}^4q+r\mbox{log}^3q)$. Notice that the algorithm  can
not run in polynomial time if $r$ is sufficiently large.

\section{Conclusion}

The basic idea of Adleman-Manders-Miller root extraction method and its complexity analysis have
not specified in the past decades. In this paper, we clarify the
method and analyze its complexity. We think our heuristic presentation is
helpful to grasp the method entirely and deeply.

 \emph{Acknowledgements.}    We thank the anonymous referees' for their detailed  suggestions.
  This work is supported by the National Natural Science
Foundation of China (Project 60873227, 11171205), and the Key Disciplines of
 Shanghai Municipality (S30104).

\end{document}